\documentclass[twocolumn,showpacs,preprintnumbers,amsmath,amssymb]{revtex4}
\usepackage{amsmath,amssymb,graphics,epsfig,subfigure}
\usepackage{color}

\begin{document}

\thispagestyle{empty}

\begin{center}

\title{Observing dynamic oscillatory behavior of triple points among black hole thermodynamic phase transitions}

\date{\today}
\author{Shao-Wen Wei$^{1,2}$ \footnote{E-mail: weishw@lzu.edu.cn, corresponding author},
        Yong-Qiang Wang$^{1,2}$,
        Yu-Xiao Liu$^{1,2}$, and Robert B. Mann$^{3}$}

 \affiliation{$^{1}$Lanzhou Center for Theoretical Physics, Key Laboratory of Theoretical Physics of Gansu Province, School of Physical Science and Technology, Lanzhou University, Lanzhou 730000, People's Republic of China\\
 $^{2}$Institute of Theoretical Physics $\&$ Research Center of Gravitation,
Lanzhou University, Lanzhou 730000, People's Republic of China\\
$^{3}$Department of Physics and Astronomy, University of Waterloo, Waterloo, Canada, N2L 3G1}

\begin{abstract}
Understanding the dynamic process of black hole thermodynamic phase transitions at a triple point is a huge challenge. In this paper, we conduct the first investigation of dynamic phase behavior at a black hole triple point. By numerically solving the Smoluchowski equation near the triple point for a six-dimensional charged Gauss-Bonnet anti-de Sitter black hole, we report that initial small, intermediate, or large black holes can transit to the other two coexistent phases at the triple point, indicating that thermodynamic phase transitions can indeed occur dynamically. More significantly, we observe characteristic weak and strong oscillatory behavior in this dynamic process, which can be understood from an investigation of the rate of first passage from one phase to another. Our results further an understanding of the dynamic process of black hole thermodynamic phase transitions.
\end{abstract}


\maketitle
\end{center}

\section{Introduction}

Black holes are now widely believed to behave as thermodynamic systems after establishing the four laws of black hole thermodynamics \cite{Hawking,Bekenstein,Bardeen}. As such, they can be expected to undergo phase transitions, one of the first being the well-known Hawking-Page phase transition \cite{Hawking:1982dh}. Over the past decade, black holes exhibit an abundance of phase behavior \cite{Kubiznak:2016qmn} on interpreting the cosmological constant in anti-de Sitter (AdS) space as the pressure of the black hole system in extended phase space \cite{Kastor,Kubiznak}. Known as black hole chemistry, small/large charged black hole phase transitions take place, reminiscent of the liquid-gas phase transition of a van der Waals (VdW) fluid \cite{Kastor,Kubiznak}.

The phase transitions including the small/large black hole phase transition have been understood using the quasi-normal modes of black holes \cite{Shen,Rao,Koutsoumbas,Myung,He,Abdalla,LiuZou,Tang,ZouLiuYue,Zangeneh,ZouZhang}. Recent studies of the dynamics of phase transitions have been conducted \cite{Li,LiWang,LiWang2} by solving a special case of the Fokker-Planck equation known as the Smoluchowski equation (SE). A dynamic transition indeed occurs between the unstable and stable black hole phases, and further investigation confirmed that even between stable small and large black hole phases, the dynamic process could occur \cite{Weiwa}. The SE has been used to analytically study the evolution of the black hole mass distribution and supermassive black hole merger rates \cite{Taniguchi,Erickcek}.

The discovery of black hole triple points \cite{Altamirano,Wei2,Frassino:2014pha} strengthens the chemical interpretation of black hole thermodynamics. It suggests that certain black hole systems are similar to water, where solid, liquid, and gas phases can coexist, with the triple point being a coexistence phase of stable small, intermediate, and large black holes. Across the triple point, the three phases can transit into one another. Consequently triple points expose significant characteristic properties of the system. However the dynamics of this transition process for black holes is still unknown.

In this study, we test and uncover such dynamic processes. In doing so, we shall discover important properties for further understanding black hole thermodynamics and, by implication, their underlying degrees of freedom.

\begin{figure}
\includegraphics[width=6cm]{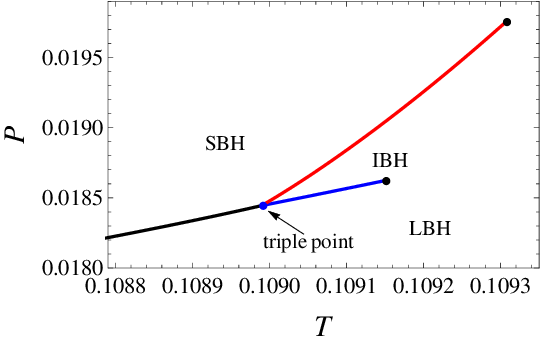}
\caption{Phase structures near the triple point for the six dimensional charged GB-AdS black holes with $Q$=0.2 and $\alpha$=1.06 in the $P$-$T$ diagram. ``SBH", ``IBH", and ``LBH" are for the small, intermediate, and large black holes.}\label{PHSTRPT}
\end{figure}

\section{Triple point in black hole systems}

To show the rich structure of the triple point, we start with the six-dimensional charged Gauss-Bonnet (GB) AdS black holes \cite{Wei2,Frassino:2014pha}. The corresponding equation of state in the extended phase space reads
\begin{equation}
 P=\frac{T}{r_{\rm h}}-\frac{3}{4\pi r_{\rm h}^2}+\frac{2\alpha T}{r_{\rm h}^3}-\frac{\alpha}{4\pi r_{\rm h}^4}+\frac{Q^2}{8\pi r_{\rm h}^8},
\end{equation}
where $P$ and $T$ are the pressure and temperature of the black hole system. All quantities are measured in Planck units, with $\alpha$ the GB coupling parameter, $Q$ denoting the black hole charge, and $r_{\rm h}$ the horizon radius, characterizing the size of the black hole as either small, intermediate, or large \cite{Kubiznak:2016qmn,Boulware,Cai2,Cvetic2,Cai}.

\begin{figure}
\includegraphics[width=6cm]{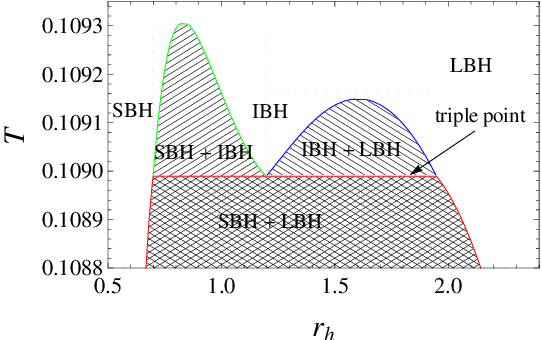}
\caption{Phase structures near the triple point in the $T$-$r_{\rm h}$ space, with $Q=0.2$ and $\alpha=1.06$.}\label{PHSTRPTr}
\end{figure}

This equation of state possesses a triple point, shown as a blue dot in Fig. \ref{PHSTRPT} in the $P$-$T$ phase diagram, where $Q$=0.2 and $\alpha$=1.06.   The other two black dots signify critical points. The diagram is similar to that of the phase structure of water. The degenerate coexistence regions in the $P$-$T$ diagram are better illustrated in a $T$-$r_{\rm h}$ phase diagram, shown in Fig. \ref{PHSTRPTr} for the first time. The two critical points are located at the two peaks, and the triple point becomes a horizontal line, along which these three black hole coexistence phases can coexist.

Consider a canonical ensemble at temperature $T_E$ (and not the Hawking temperature), composed of a series of black hole spacetimes (called the landscape) with arbitrary horizon radius, and define  a new Gibbs free energy
\begin{align}
 &G_L = H-T_{\rm E}S  \label{glll}\\
&=\frac{2\pi r^5_{\rm h}}{15}\left(4\pi P - \frac{5\pi T_{\rm E}}
   {r_{\rm h}}+ \frac{5}{r_{\rm h}^2} - \frac{20\pi\alpha T_{\rm E}}{r_{\rm h}^3} +\frac{5\alpha}{r_{\rm h}^4}\right)+\frac{\pi Q^2}{9r_{\rm h}^3} \nonumber
\end{align}
where $H$ and $S$ are the respective enthalpy and entropy. This generalized off-shell free energy describes transient black hole states with $r_{\rm h}$ as the order parameter, describing the underlying microscopic degrees of freedom \cite{LiWang}.

\begin{figure}
\includegraphics[width=6cm]{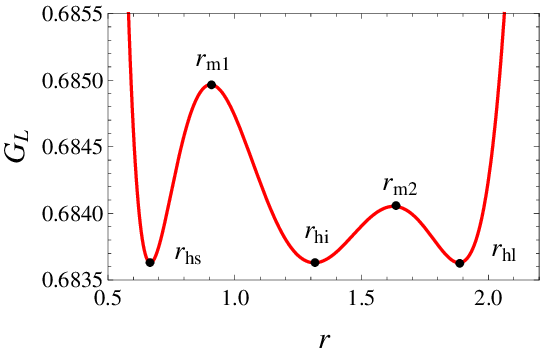}
\caption{Behavior of the Gibbs free energy via landscape at the triple point with $Q$=0.2, $T_{\rm E}$=0.108, and $P$=0.018164.}\label{GibbsLSa}
\end{figure}

We illustrate $G_L$ at the triple point in Fig. \ref{GibbsLSa} for an ensemble temperature $T_{\rm E}$=0.108 with $Q$=0.2, $P$=0.018164, $\alpha$=1.080978, $r_{\rm hs}$=0.667162, $r_{\rm hi}$=1.313823, $r_{\rm hl}$=1.884450, $r_{\rm m1}$=0.905109, and $r_{\rm m2}$=1.626470. Each point on this curve denotes a black hole state; however, not all of them are actual black hole solutions of the Einstein GB equations. Only the local extrema correspond to actual black holes, while the others are off-shell transient states \cite{Weiwa}. The local minima and maxima denote thermodynamic stable or unstable black holes. Combining the fact that a stable thermodynamic system has the lowest free energy, the black hole system will migrate to (or remain at) the well of the deepest depth. For the parameter choice employed in Fig. \ref{GibbsLSa}, the three wells have the same depth, indicating a triple point where the black hole system can be in any arbitrary combination of these three states.

\section{Dynamic processes at the triple point}

Based on the behavior of the Gibbs free energy $G_L$, we examine the dynamic properties of black hole phase transitions at the triple point. Given an initial
state in the black hole system (e.g. a small black hole), at the triple point it will evolve via a diffusion process to other states because it can undergo thermodynamic phase transitions. We denote $\rho(r_{\rm h}, t)$ as the probability distribution of the system staying at a given black hole state. The SE \cite{Zwanzig}
\begin{eqnarray}
 \frac{\partial \rho(r, t)}{\partial t}=D
 \frac{\partial}{\partial r}\left(e^{-\frac{\beta G_{\rm L}(r)}{k_{\rm B}T_{\rm E}}}
 \frac{\partial}{\partial r}\left(e^{\frac{\beta G_{\rm L}(r)}{k_{\rm B}T_{\rm E}}}\rho(r, t)\right)\right) \label{FPE}
\end{eqnarray}
governs this behavior, where we have replaced $r_{\rm h}$ with $r$ for simplicity. The diffusion coefficient $D=k_{\rm B}T_{\rm E}/\zeta$ with $k_{\rm B}$ and $\zeta$ being the Boltzmann constant and dissipation coefficient. Without loss of generality, we set $k_{\rm B}$=$\zeta$=1 in the following, and choose the initial state to be a Gaussian wave packet
\begin{eqnarray}\label{rhoinit}
 \rho(r, 0)=\frac{1}{0.005\sqrt{\pi}}e^{-\frac{(r-r_{\rm \textsf{j} })^2}{0.005^2}}
\end{eqnarray}
located at $r_{\rm \textsf{j} }$. Setting $r_{\rm \textsf{j} }$=$r_{\rm hs}$, $r_{\rm hi}$, or $r_{\rm hl}$ means the initial state is peaked at a coexistent small, intermediate, or large black hole state. By imposing reflective boundary conditions at $r=0$ and $r=\infty$ (sufficiently large distance), we numerically solve SE equation for each case, illustrating the results in Fig. \ref{pprhotrc}. Regardless of the choice of initial state, we observe that the probability $\rho(r, t)$ leaks to other states, indicating that there are indeed phase transitions between these three coexistent states. This occurs very quickly: within $t=80$ the system attains its final stationary state, with the local maxima of $\rho(r, t)$ at the location of these three coexistent black holes. More detailed study indicates that these maxima share the same value $\rho(r_{\rm \textsf{k}}, t)$=0.4801, for $\textsf{k} =$ hs, hi, and hl, indicating that the black hole system settles into a combination of these coexistent phases. Moreover, this specific value is independent of the initial state, which can be understood from the SE. After a sufficiently long time, the system approaches stationarity; $\rho(r, t)$ will no longer change, and so the left side of (\ref{FPE}) vanishes. The probability distribution is then determined only by $G_L$ in (\ref{glll}).

This dynamic phase transition process at the triple point is different from that of a VdW type phase transition \cite{LiWang,LiWang2,Weiwa}. We now examine the detailed evolution of each coexistent black hole phase by plotting in Fig. \ref{pprhotc} the behavior of $\rho(r_{\rm \textsf{k}}, t)$, when the initial Gaussian wave packet peaks at these states. We see that the initially large value $\rho(r_{\rm \textsf{k}}, t)$ in each case rapidly decays to a stationary value, while the other two states (initially zero) grow toward this value.

Considering first Fig. \ref{rhota}, the initial Gaussian wave packet $\rho(r, 0)$ peaks at the small black hole phase with $\rho(r_{\rm hs}, 0)$ large, while $\rho(r_{\rm hi}, 0)$ and $\rho(r_{\rm hl}, 0)$ are negligibly small. As $t$ increases $\rho(r_{\rm hs}, t)$ decreases, and $\rho(r_{\rm hi}, t)$ and $\rho(r_{\rm hl}, t)$ increase as expected, the latter more slowly because the initial state must surmount two barriers. The transition rate from the small to intermediate black hole state is higher than the total rate from intermediate to small and large black hole states. Once $t \geq 40$, all of them tend to 0.4801, where the final stationary state is achieved. During the evolution, we observe an interesting phenomenon for $\rho(r_{\rm hi}, t)$: it then increases to a maximum of 0.5338 at $t=3.6986$ and then decreases with time to its stationary value of 0.4801. This presumably occurs because the intermediate black hole state can transit to both small and large black hole states, while the system will stay longer at the intermediate black hole state than the small and large black hole states, see the mean first passage times obtained in the next section. As the large black hole state becomes more populated, it will transit back to the intermediate one, reducing the rate of the latter. Because $\rho(r_{\rm hi}, t) < \rho(r_{\rm hs}, t)$, we refer to this novel behavior as a weak oscillatory phenomenon.

\begin{widetext}
\begin{center}
\begin{figure}
\center{\subfigure[]{\label{Prhotra}
\includegraphics[width=4.5cm]{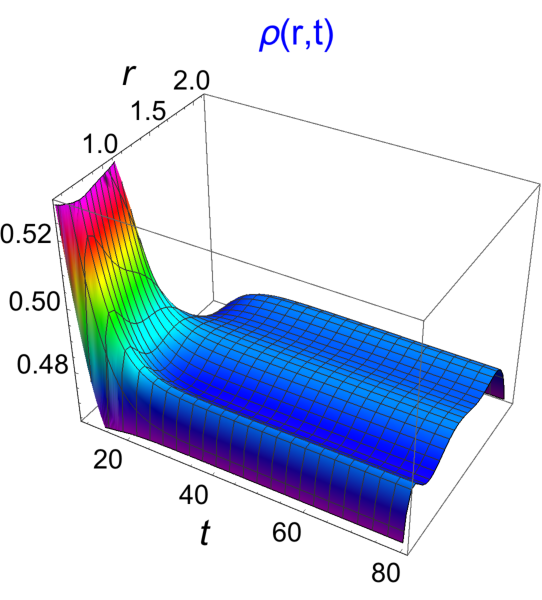}}
\subfigure[]{\label{rhotrb}
\includegraphics[width=4.5cm]{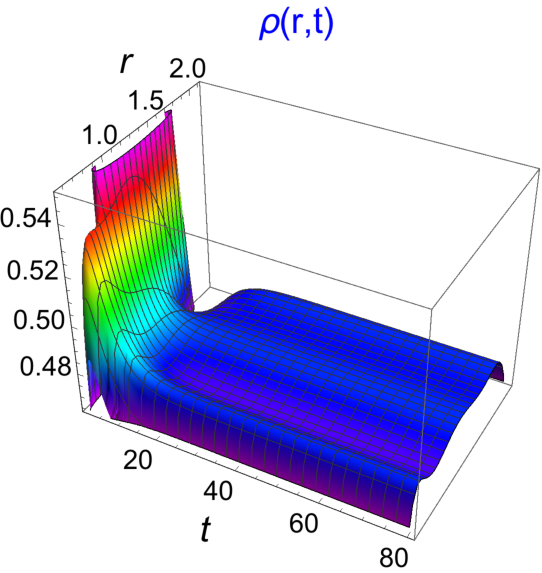}}
\subfigure[]{\label{rhotrc}
\includegraphics[width=4.5cm]{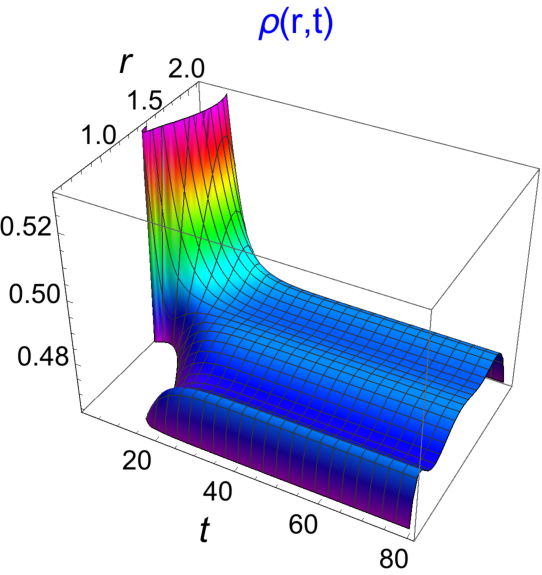}}}
\caption{Probability distribution $\rho(r, t)$ governed by the SE equation with $Q$=0.2, $T_{\rm E}$=0.108, and $P$=0.018164. The initial Gaussian wave packet, respectively, located at (a) small black hole state, (b) intermediate black hole state, and (c) large black hole state.}\label{pprhotrc}
\end{figure}
\begin{figure}
\center{\subfigure[]{\label{rhota}
\includegraphics[width=4.5cm]{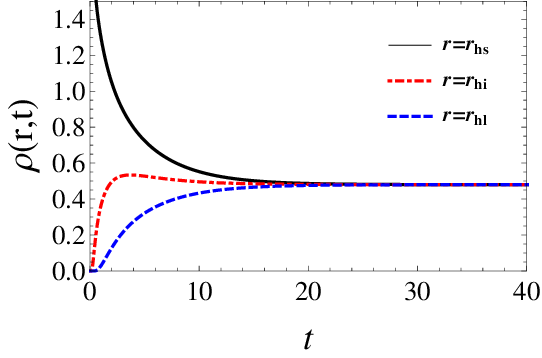}}
\subfigure[]{\label{rhotb}
\includegraphics[width=4.5cm]{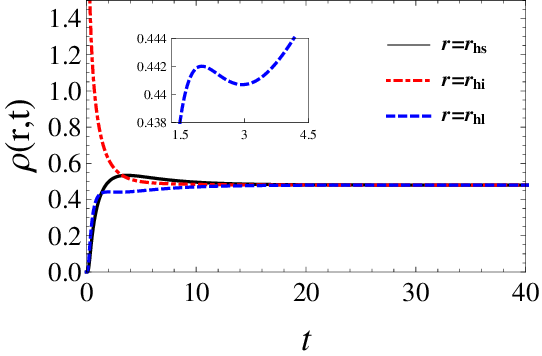}}
\subfigure[]{\label{rhotc}
\includegraphics[width=4.5cm]{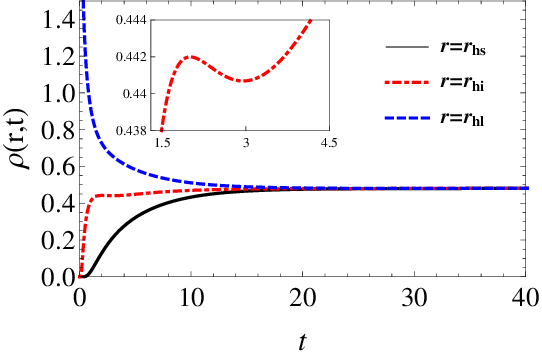}}}
\caption{Behaviors of the probability $\rho(r, t)$ at the coexistence small, intermediate, and large black hole states, when the initial Gaussian wave packet is peaked at the coexistent (a) small (b) intermediate and (c) large black hole states.}\label{pprhotc}
\end{figure}
\end{center}
\end{widetext}

Corresponding behavior is observed when the initial wave packet is peaked at the intermediate black hole state, shown in Fig. \ref{rhotb}. Here, a more fine structure is present. As expected, $\rho(r_{\rm hi}, t)$ decreases whereas $\rho(r_{\rm hs}, t)$ and $\rho(r_{\rm hl}, t)$ both increase, with the former reaching a maximum of 0.5337 at $t=3.6987$. At early times, these probability distributions exhibit interesting behavior. When $0<t<1.3840$, we find that the probability leaks from the intermediate to both small and large black hole states, with $\rho(r_{\rm hl}, t)$ increasing faster than $\rho(r_{\rm hs}, t)$ because of the lower barrier height. This pattern reverses for $t>1.3840$. Once $t>3.2995$, $\rho(r_{\rm hs}, t)$ is dominant among the three probabilities, indicating that the system has a large probability of staying at the coexistent small black hole state. For a short time afterward $\rho(r_{\rm hs}, t)$ continues to grow to its maximum, after which it decays to stationarity, transiting back to the coexistent intermediate and large black hole states. Since $\rho(r_{\rm hs}, t)$ eventually dominates we refer to this as a strong oscillatory phenomenon. In Fig.~\ref{rhotc}, we see that for all $t$, $\rho(r_{\rm hl}, t)>\rho(r_{\rm hi}, t)>\rho(r_{\rm hs}, t)$. However, we observe additional (tiny) oscillatory behavior (shown in the insets in Fig. \ref{rhotb} and \ref{rhotc}) when the initial wavepacket is initially peaked at the coexistent intermediate or large black hole phases.

\section{First passage event}

We seek here more clues about these oscillatory phenomena from the distribution of first passage times of the phase transition. This is defined as the rate at which a given initial state first reaches an unstable black hole phase, represented by the peak of the free energy \cite{LiWang}. From Fig.~\ref{GibbsLSa}, we have three cases: small to intermediate (case I), intermediate to small and large (case II), and large to intermediate (case III). For all cases, we impose reflective boundary condition at $r=0$ and $r=\infty$ (sufficiently large $r$) and absorbing boundary conditions at the peaks $r_{\rm m1}$ and $r_{\rm m2}$, accordingly, where the latter model a given coexistent state first leaving the system. Using the SE, we can express the first passage rate $F_{P}(t)$ for these three cases as
\begin{eqnarray}
 F_{P1}(t)&=&-\frac{\partial \rho(r_{\rm m1}, t)}{\partial r},\\
 F_{P2}(t)&=&\frac{\partial \rho(r_{\rm m1}, t)}{\partial r}-\frac{\partial \rho(r_{\rm m2}, t)}{\partial r},\\
 F_{P3}(t)&=&\frac{\partial \rho(r_{\rm m2}, t)}{\partial r}.
\end{eqnarray}

\begin{figure}
\center{\subfigure[]{\label{Fpt}
\includegraphics[width=4.1cm]{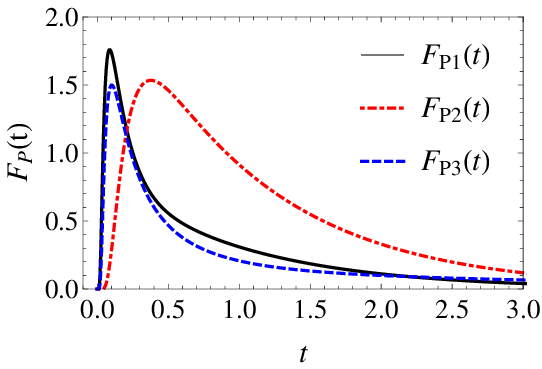}}
\subfigure[]{\label{Fp2t}
\includegraphics[width=4.1cm]{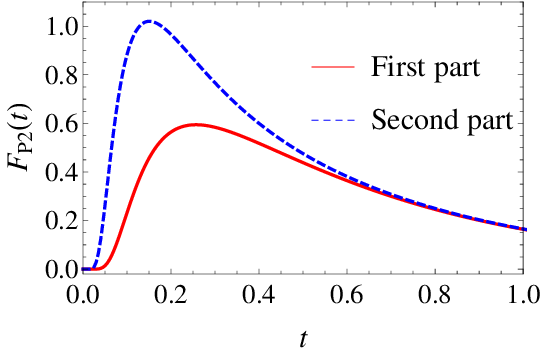}}}
\caption{(a) First passage time rate for case I, II, and III. (b) The first and second parts of the first passage time for $F_{P2}(t)$.}\label{ppFp2t}
\end{figure}

The numerical results are given in Fig. \ref{Fpt}. For each case, there is a single peak at $t_1$=0.0876, $t_2$=2.3725, and $t_3$=0.1026, which can be interpreted as the length of time at which the system remains in its initial state for each case before first transiting to another state. For cases I and III, the system first transits to other states after a short time, whereas for case II, the system remains in the intermediate black hole state somewhat longer. This expectation is borne out from the mean first passage time
\begin{equation}
 \langle t\rangle=\int_{0}^{\infty}tF_{P}dt,\label{mfpt}
\end{equation}
for each case, where we find $\langle t_1\rangle$=0.86, $\langle t_2\rangle$=2.37, $\langle t_3\rangle$=1.75 (where $0<t<40$).

For case I, the initial state is at the coexistent small black hole state. As time increases, its probability $\rho(r_{\rm hs}, t)$ decreases and leaks to the intermediate black hole state. Since $\langle t_2\rangle$ is larger than $\langle t_1\rangle$ and $\langle t_3\rangle$, the system will stay longer at the intermediate black hole state. Then, it will transit back to the coexistent small or large black hole states. Consequently, $\rho(r_{\rm hi}, t)$ first increases then decreases with time, and thus   weak oscillatory behavior is present in Fig. \ref{rhota}.

We report that $F_{P}(t)$ is governed not only by the barrier height, as discussed previously \cite{Li,LiWang}, but also by the barrier width. An inspection of Fig.~\ref{GibbsLSa} indicates that the difference in barrier heights $\Delta G_L = G_{m1} - G_{m2} \simeq0.001$, whereas the difference in barrier widths is $\Delta r_{m2}$-$\Delta r_{m1}\simeq$0.02, 20 times larger than $\Delta G_L$. These larger barrier width values will dominate over the small barrier height difference, with smaller values of $\Delta r$, leading to a peak in $F_{P}(t)$ at small $t$. Since $\Delta r_{m1}=0.2379$ (case I) is slightly smaller than $\Delta r_{m2}=0.2580$ (case III), we have $t_1<t_3$. For case II absorbing boundary conditions are imposed at both sides; therefore the intermediate black hole can transit to large and small black holes through the right and left boundaries.

To understand the strong oscillatory behavior shown in Fig. \ref{rhotb}, we plot the two parts of $F_{P2}(t)$ with red solid and blue dashed curves in Fig. \ref{Fp2t}. The first and second parts denote the transitions to the small and large black hole states. Their peaks are at $t_{i1}$=0.2568 and $t_{i2}$=0.1507, which is consistent with the fact that the right barrier is lower and closer to the well than the left one. However, we find from (\ref{mfpt}) that $\langle t_{i1}\rangle =0.28< \langle t_{i2}\rangle =0.31$. Hence $\rho(r_{\rm hs}, t)$ will at some time become larger than $\rho(r_{\rm hl}, t)$. Thus, strong oscillatory behavior is exhibited when the initial state is peaked at the coexistent intermediate black hole state. The fact that $t_{i1}>t_{i2}$ indicates that $\rho(r_{\rm hl}, t)$ increases faster than $\rho(r_{\rm hs}, t)$ at early transition times. It is worth pointing out that the location of the peak of $F_{P}(t)$ has no direct relationship to the mean first passage time.

\section{Summary}

We have conducted the first investigation of dynamic phase behavior at a black hole triple point. The degenerate triple point in
a $P$-$T$ phase diagram becomes a horizontal line in the $T$-$r_{\rm h}$ phase diagram, with each coexistence region explicitly displayed. On the free energy landscape, with $r_{\rm h}$ as the order parameter, the characteristic pattern of the triple point is that of three potential wells of the same depth, with the coexistent black hole phases located at the minima of the wells.

Transitions between these different phases are different from that of first-order VdW type phase transitions, and can be studied dynamically via the SE. We found that, regardless of the initial phase in the black hole system, each phase can transit into the others, attaining stationarity of all three phases after a short time. More significantly, we observe both weak oscillatory behavior, in which the probability distribution of one state attains a maximal value before decaying to stationarity, and strong oscillatory behavior, in which this maximum dominates over the other two states. Such behaviors are also well understood by calculating the mean first passage time.

Our results indicate that black holes have interesting dynamic phase behavior whose relationship to their underlying degrees of freedom remains to be understood. It would be interesting to apply our approach to other black hole systems with various types of phase transitions, including angular momenta \cite{Altamirano:2014tva}, hair \cite{Giribet,Hennigar,Dykaar:2017mba}, and acceleration \cite{Anabalon:2018ydc,Anabalon:2018qfv}, where more interesting dynamic phenomena should be observed.

{\emph{Acknowledgements}.}---This work was supported by the National Natural Science Foundation of China (Grants No. 12075103, No. 11675064, No. 11875151, and No. 12047501) and the Natural Sciences and Engineering Research Council of Canada.

\end{document}